# R$_1$ dispersion contrast at high field with fast field-cycling MRI

Markus Bödenler[a], Martina Basini[b], Maria Francesca Casula[c], Evrim Umut[d], Christian Gösweiner[a], Andreas Petrovic[a], Danuta Kruk[d], Hermann Scharfetter[a]

[a]*Institute of Medical Engineering, Graz University of Technology, Stremayrgasse 16, A-8010 Graz, Austria*
[b]*Physic Deppartment and INSTM, Università degli Studi di Milano, Via Celoria 16, I-20133 Milan, Italy*
[c]*Department of Chemical and Soil Sciences and INSTM, University of Cagliari, I-09042 Monserrato (CA), Italy*
[d]*Faculty of Mathematics and Computer Science, University of Warmia & Mazury, Słoneczna 54, 10-710 Olsztyn, Poland*

**Corresponding author:** Markus Bödenler, m.boedenler@tugraz.at

## Highlights

- First FFC-MRI hardware setup for a clinical field strength of 3 T
- Iron oxide magnetic nanoparticles proved as suitable FFC-MRI contrast agent
- Proof-of-principle of delta relaxation enhanced MR imaging at 3 T

## Abstract

Contrast agents with a strong R$_1$ dispersion have been shown to be effective in generating target-specific contrast in MRI. The utilization of this R$_1$ field dependence requires the adaptation of a MRI scanner for fast field-cycling (FFC). Here, we present the first implementation and validation of FFC-MRI at a clinical field strength of 3 T. A field-cycling range of ±100 mT around the nominal B$_0$ field was realized by inserting an additional insert coil into an otherwise conventional MRI system. System validation was successfully performed with selected iron oxide magnetic nanoparticles and comparison to FFC-NMR relaxometry measurements. Furthermore, we show proof-of-principle R$_1$ dispersion imaging and demonstrate the capability of generating R$_1$ dispersion contrast at high field with suppressed background signal. With the presented ready-to-use hardware setup it is possible to investigate MRI contrast agents with a strong R$_1$ dispersion at a field strength of 3 T.

## Keywords

## Graphical abstract

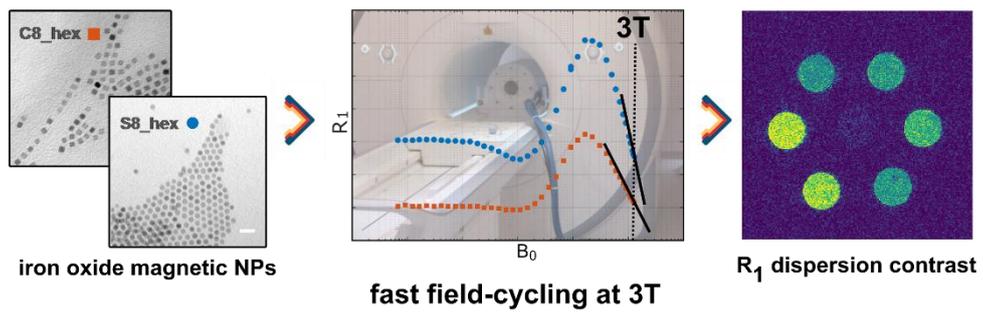

## 1. Introduction

As the name implies, fast field-cycling magnetic resonance imaging (FFC-MRI) is the combination of fast field-cycling nuclear magnetic resonance (FFC-NMR) relaxometry with imaging methods of MRI and is based on cycling the $B_0$ field within an imaging sequence [1]. This gives access to new types of contrasts arising from the field dependency of the $^1$H relaxation rates $R_1$ and $R_2$, also termed as $R_1$ and $R_2$ nuclear magnetic relaxation dispersion (NMRD), respectively. FFC-MRI can be implemented on clinical MRI systems by inserting an additional $B_0$ insert coil into an otherwise conventional scanner. This was previously realized for a clinical field strength of 1.5 T and is also referred to as delta relaxation enhanced magnetic resonance (dreMR) imaging [2–5].

Alford et al. [2] demonstrated a differentiation between bound and unbound contrast agent (CA) using the targetable probe gadofosveset which only exhibits a strong $R_1$ dispersion upon binding to serum albumin. This leads to an improved target specificity by exploiting image contrast based on the increased $dR_1/dB_0$ of the bound CA instead of the absolute difference of the relaxation rate at a fixed magnetic field. Therefore, the dreMR image shows only contrast arising from the bound CA and supresses contrast from the anatomical background as well as from the unbound agent, which has been recently shown by in-vivo experiments with mice [6]. Hoelscher et al. [4] extended the dreMR theory to quantitative concentration measurements using a correction for finite ramp times during field-cycling and a compensation of field-cycling induced eddy current fields by dynamic reference phase modulation [7].

Whereas the $R_1$ dispersion of healthy tissue is large for low magnetic fields, in the range of clinical field strengths, the dispersion is inherently weak (e.g. -0.19 s$^{-1}$T$^{-1}$ around 1.5 T for murine muscle tissue), as recent findings by Araya et al. [6] have emphasized. Therefore, the use of contrast agents exhibiting a strong dependence of $R_1$ upon the magnetic field i.e. a steep slope $dR_1/dB_0$ in the NMRD profile is favourable to obtain significant $R_1$ dispersion contrast.

Up to now, all suitable CAs for dreMR utilize a preferably steep slope in the NMRD profile. An interesting alternative for dreMR imaging at clinical fields could be the exploitation of extrinsic contrast agents based on quadrupole relaxation enhancement (QRE). High spin quadrupole nuclei (QN) such as $^{209}$Bi offer the potential of QRE peaks emerging in the clinical $B_0$ range instead of a smooth dispersion [8].

The cross-relaxation between water protons and quadrupole nuclei, i.e. nuclei with a spin quantum number > 1/2, gives rise to a shortening of the longitudinal proton relaxation time $T_1$. This so called quadrupole relaxation enhancement [9–11] offers a high potential for designing smart molecular probes for the usage as MRI contrast agents in the context of cellular and molecular imaging [8]. So far the effectiveness of QRE for increasing MRI contrast has been shown at low magnetic fields for the cross-relaxation between $^{14}$N and $^1$H in protein backbones [1,12,13]. However, this effect is entirely unexplored for the design of extrinsic contrast agents at clinical field strengths such as 1.5 T or 3 T. QRE based CAs are frequency selective as the cross-relaxation can only become effective if the proton Larmor frequency matches one of the transition frequencies of the QN [11]. This favourable feature offers the possibility to activate and inactivate QRE, and therefore image contrast, by modulating the magnetic field. The frequency position of QRE can either be altered by chemical interaction with the biological environment allowing for chemically selective contrasts or by shifting of the main $B_0$ field of the MRI system itself, namely FFC-MRI.

The validation and future application of QRE contrast agents at clinical fields requires a dedicated MRI system adapted for fast field-cycling. Although there exist a handful of 1.5 T systems worldwide, we show, to the best of our knowledge, the first FFC-MRI setup for a field strength of 3 T. To this end, the aim of this work is to present important steps for the implementation and validation of a small animal FFC-MRI system for 3 T by means of a $B_0$ insert coil. We describe the specification of our hardware setup and show a rather simple approach to overcome imaging artefacts due to random phase fluctuations induced by the output noise current of the gradient power amplifier. Furthermore, system validation is performed by FFC-NMR relaxometry measurements and proof-of-concept dreMR imaging of selected iron oxide magnetic nanoparticles (IOMNP). The final result is a ready-to-use FFC-MRI system for a clinical field strength of 3 T with the envisaged aim of exploiting the magnetic field dependency of MRI contrast agents in general and to investigate the imaging potential of prospective QRE compounds in particular.

## 2. Methods

### Hardware setup

The field-cycling hardware was implemented on a clinical 3 T MRI system (Skyra, Siemens Healthineers, Germany) using a custom-built $B_0$ insert coil (Resonance Research Inc., USA). The insert coil has a resistance of 63.8 mΩ and an inductance of 1.69 mH. It is driven by a gradient power amplifier (IECO, Finland) capable of a maximum continuous output current of ±150 A (±300 A peak) and a field efficiency of 0.668 mT/A. This allows for an offset field $\Delta B_0$ of ±100 mT within a minimum ramp time of 1 ms. Field inhomogeneities are less than 1 % over a diameter of 35 mm in axial direction. Shielded design of the coil reduces the fringe field to 0.2 mT at the main magnet bore (radius of 35 cm). The insert is placed in the isocenter of the main magnet and holds a custom-built transmit/receive [1]H birdcage radio frequency (RF) coil (MRI.TOOLS GmbH, Germany) with variable tune and match capability. This enables an imaging region of 40 mm in axial and 40 mm in longitudinal direction, which is suitable for imaging of phantoms and small animals. The $\Delta B_0$ pulses are generated with an arbitrary waveform generator controlled by an optical trigger signal from the MRI pulse sequence. This design ensures that there is no electronic coupling to the main magnet system.

A basic FFC-MRI pulse sequence consists of a magnetization preparation such as an inversion (IR) or a saturation (SR) RF pulse followed by an evolution phase of duration $T_{evol}$ with applied field shift $B_0 \pm \Delta B_0$ and subsequent signal acquisition at the nominal $B_0$ field. The trigger defines the timing of the $\Delta B_0$ pulse within the pulse sequence synchronizing the $B_0$ insert coil with the MRI system. A schematic of the aforementioned timing is illustrated in Figure 1 and thorough reviews of pulse sequences can be found in [4,5,14,15].

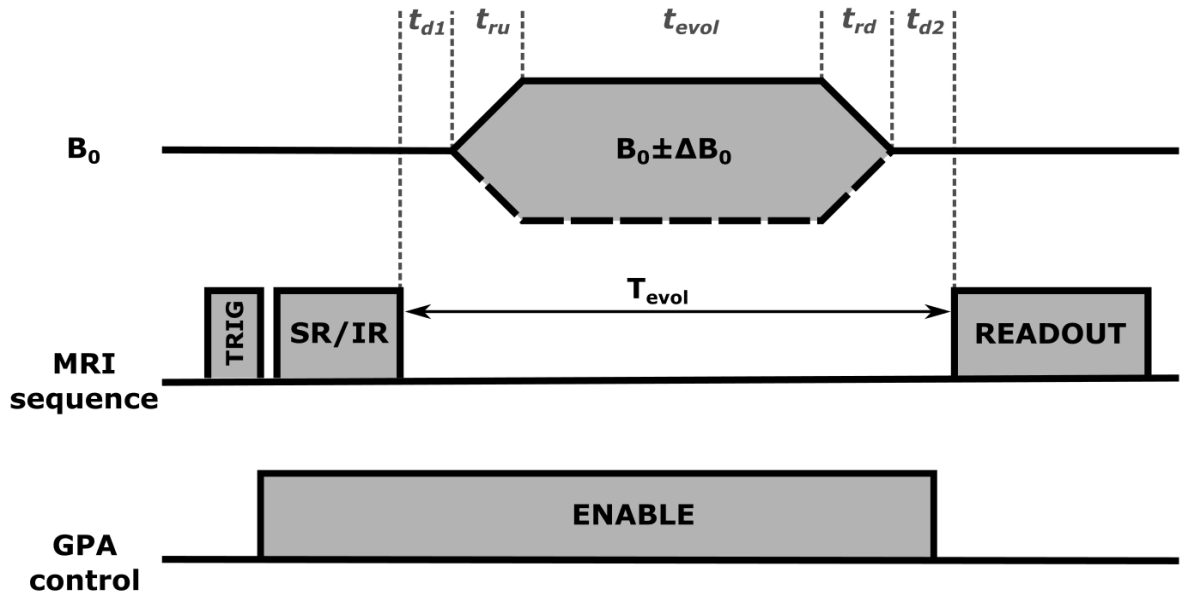

Figure 1: Schematic of a FFC-MRI pulse sequence with saturation (SR) or inversion (IR) magnetization preparation. The trigger (TRIG) synchronizes the $B_0$ insert coil with the MRI system. The ENABLE signal is used for blanking the GPA during image readout to prevent imaging artefacts due to random phase fluctuations. The timing of the $\Delta B_0$ pulse is as follows: delay time after the preparation phase $t_{d1}$, ramp up and ramp down times $t_{ru}$ and $t_{rd}$, respectively, flat top time $t_{evol}$ and settling time $t_{d2}$. $T_{evol}$ denotes the total time between magnetization preparation and image acquisition.

Whereas magnetic field inhomogeneities in the order of 1 % during the evolution phase are not critical for FFC-MRI, the field stability during signal acquisition is crucial to prevent phase errors. In our case the output noise current of the gradient power amplifier (GPA) in the idle state induces random phase fluctuations during image acquisition causing serious ghosting artefacts in phase encoding direction. This problem has also been observed in other FFC-MRI systems [5,16,17]. Recently, a post-processing correction technique was proposed minimizing the background signal by determining the optimum phase correction for each line in k-space [17]. Another hardware based approach is to alternate the amplifier load between the $B_0$ insert coil and an inductance-matched dummy load using a fast high-power solid state switch [5].

Here, we choose an approach utilizing an additional synchronized TTL signal to the dedicated control input of the GPA. The GPA is only enabled during the evolution phase and disabled throughout image acquisition (see ENABLE signal in Figure 1) so as to suppress the output noise current sufficiently. There is, however, a timing restriction in the hardware of the GPA: According to the specifications there is a delay between the edges of the control signal for enable and disable and the physical response of the GPA output stage, i.e. about 20 ms (enable) and 2 ms (disable), respectively. Consequently, the trigger signal has to be sent earliest 20 ms prior to the rising edge of the actual $\Delta B_0$ pulse to ensure a safe operation mode of the GPA.

To show the effectiveness of this approach, images were acquired with and without GPA blanking using a modified saturation recovery spin echo (SR-SE) sequence with the following parameters: repetition time TR = 1000 ms, echo time TE = 15 ms, $T_{evol}$ = 150 ms, field of view

FOV = 50 mm x 50 mm, matrix size 256 x 256, bandwidth BW = 130 Hz/Px, flip angle = 90° and slice thickness = 5 mm. No field shift was applied for these measurements.

Eddy current characterization and compensation

Eddy current related artefacts are a key challenge for FFC-MRI [5,7], in particular, when two images with contrast from different $B_0$ field shifts are subtracted. Thus, the characterization and compensation of the spurious signal dynamics induced by the eddy current field is crucial. Fast ramping of the offset field induces eddy currents in the main magnet structure of the MRI system. Though the $B_0$ insert coil is actively shielded to reduce the coupling to the main magnet system, the residual fringe field results in a temporally varying offset of the Larmor frequency. This offset gives rise to a rigid-body image shift in frequency and slice encoding direction.

The temporal behaviour of the eddy currents (i.e. the Larmor frequency shift) is characterized using a modified multi-phase spoiled gradient echo sequence [18]. A phantom holding six samples with different concentrations (two-fold serial dilution beginning with 1 mM) of the contrast agent Gadovist® (Bayer Vital GmbH, Germany) and a water reference sample was prepared. After an applied $\Delta B_0$ pulse of +100 mT and $T_{evol}$ of 300 ms a total number of 50 images (phases) were acquired with a repetition interval of 17 ms, resulting in a sampling duration of 850 ms during the eddy current decay. The remaining imaging parameters were as follows: TE = 8.5ms, FOV = 50 mm x 50 mm, matrix size = 64 x 64, BW = 100 Hz/Px, flip angle = 15° and slice thickness = 5 mm. The $\Delta B_0$ was cycled with a repetition time of 3000 ms. Mono- and bi-exponential decay curves were fitted to the obtained frequency shift resulting in an accurate model of the eddy current decay.

During the acquisition of FFC-MRI images, eddy currents then are compensated by dynamic reference phase modulation (eDREAM) as proposed in [7]. This method can be directly implemented into the MRI pulse sequence and does not require any additional hardware adaptations. To validate the eDREAM implementation a reference image (without applied field shift) and an image at shifted $B_0$ ($\Delta B_0$ of +100 mT and $T_{evol}$ of 300 ms) of the aforementioned Gadovist® phantom were acquired, respectively, using a modified SR-SE sequence with parameters: TR/TE = 3000/15 ms, FOV = 50 mm x 50 mm, matrix size = 256 x 256, BW = 130 Hz/Px, flip angle = 90° and slice thickness = 5mm. The shifted image was normalized to account for the different equilibrium magnetizations [4] and the difference image was obtained by magnitude subtraction of the reference image and shifted image.

Dispersive contrast agent for 3 T and FFC measurements

In order to validate the implementation of the FFC-MRI hardware setup, the availability of a contrast agent exhibiting strong $R_1$ relaxation dispersion at 3 T is crucial. Recently published iron oxide magnetic nanoparticles [19] were selected for system validation because of a suspected $R_1$ dispersion at 3 T. Such IOMNPs are obtained by a high temperature surfactant-assisted chemical route which leads to size- and shape- controlled nanocrystals. IOMNPs with similar average size (nearly 8 nm) and with different shapes (spherical and cubical) were used dispersed in hexane, and will be referred to as S8_hex and C8_hex, respectively. Samples containing three different concentrations (1 mM, 0.5 mM and 0.25 mM) for each shape and a hexane reference were prepared and arranged as illustrated in Figure 6a. The samples were investigated by Transmission electron microscopy (TEM) on an H-7000 Microscope (Hitachi, Japan) equipped with a W gun operating at 125 kV. Fe concentration in the dispersion was assessed by quantitative analysis through inductively coupled plasma atomic emission spectrometry using a Varian Liberty 200 ICP-AES (Agilent Technologies, USA).

$^1$H $R_1$ NMRD measurements were performed for the 1 mM samples and the hexane reference sample by using a Spinmaster FFC 1 T relaxometer (Stelar, Italy) for a Larmor frequency range of 10 kHz-30 MHz. An external HTC-110 3 T superconducting magnet equipped with a standalone PC-NMR console (Stelar, Italy) was used for a frequency range of 40 MHz-128 MHz. For measurements between 10 kHz-40 MHz one-pulse FFC sequences were used: Below 10 MHz the sample is pre-polarized at a field of 0.57 T and then the magnetic field is cycled to a different value, where the spin system is allowed to relax for a variable time period. Finally, the field is cycled back to the detection value, where the FID is recorded following the application of a 90° RF pulse. Above 10 MHz, where the signal is reasonably higher than at low field, the same procedure was followed except pre-polarization was omitted. For measurements between 40-128 MHz a standard inversion recovery sequence was used. Furthermore, additional points were measured at 119 MHz, 123.2 MHz and 127.4 MHz corresponding to the achievable field cycling range (2.89 T ± 100 mT) of the $B_0$ insert coil.

Images were acquired with the FFC-MRI system at three different field strengths (2.79 T, 2.89 T and 2.99 T) and various evolution times ($T_{evol}$ = 60, 100, 150, 300, 500, 800, 1500 and 3000 ms) using a modified SR-SE sequence. The remaining imaging parameters were: TR/TE = 10000/15 ms, FOV = 40 mm x 40 mm, matrix size = 64 x 64, BW = 130 Hz/Px, flip angle = 90° and slice thickness = 5 mm. For each field strength, $R_1$ maps were estimated on a pixel-by-pixel basis by nonlinear fitting to a mono-exponential saturation recovery model [20]. In addition, a $\Delta R_1/\Delta B_0$ map was calculated by subtracting $R_1$ maps obtained for 2.99 T and 2.79 T.

DreMR signal simulation

In general, the behaviour of the magnetization in presence of an external magnetic field can be described by the Bloch equations. In the case of field-cycling, the main magnetic field $B_0$ is modulated throughout the pulse sequence and thus is a function of time. As the equilibrium magnetization $M_0$ and the longitudinal relaxation rate $R_1$ depend on $B_0$, they exhibit also a time-dependent behaviour requiring a numerical solver for the Bloch equations. However, under several assumptions for the waveform of the $\Delta B_0$ pulse, closed-form solutions exist for the longitudinal magnetization $M_z$ [12,15]. To predict the behaviour of $M_z$ as a function of the evolution time and for different field shifts we slightly adapted the proposed equation in [15] for saturation recovery and an additional delay time before the applied field shift. The common used linear ramps can be approximated by a step response with relaxation for half of the time at the nominal field and the other half at the desired field shift. The change in $M_0$ and $R_1$ can be assumed to be linear with respect to $\Delta B_0$ within the field-cycling range. This gives a full description of the magnetization behaviour $M_z^\pm$ under consideration of the underlying timing restrictions in a FFC-MRI pulse sequence:

$$M_z^\pm(t_{evol}) = M_0 + \left\{M_0^\pm - M_0 - \left\{M_0^\pm - M_0 \right.\right.$$
$$\left.\left. + \alpha M_0 \exp\left[-\left(\frac{t_{ru}}{2} + t_{d1}\right)R_1\right]\right\} \exp\left[-\left(t_{evol} + \frac{t_{ru} + t_{rd}}{2}\right)R_1^\pm\right]\right\} \exp\left[-\left(\frac{t_{rd}}{2} + t_{d2}\right)R_1\right] \quad (1)$$

$$\text{where} \quad M_0^\pm = M_0\left(1 \pm \frac{\Delta B_0}{B_0}\right). \quad (2)$$

$M_0$ and $M_0^\pm$ are the equilibrium magnetizations, $R_1$ and $R_1^\pm$ the relaxation rates at nominal and shifted $B_0$ field, respectively. The superscript $\pm$ indicates whether a positive or negative field shift is applied. For saturation recovery the parameter α should be equal to 1, $t_{d1}$ is the delay time between saturation pulse and begin of the ramp up time $t_{ru}$, $t_{rd}$ is the ramp down time, $t_{evol}$ is the flat top time and $t_{d2}$ is the settling time after the $\Delta B_0$ pulse. $T_{evol}$ is given by the total time between magnetization preparation and begin of the image acquisition. This timing of $\Delta B_0$ within the pulse sequence is illustrated in Figure 1. Exemplarily, we simulated the relaxation curves for a positive and negative field shift of ±100 mT for the sample containing 0.5 mM of cubic-shaped IOMNPs (ROI 2 in Figure 6a). The estimated $R_1$ values as well as $M_0$ from the aforementioned measurements were used for the simulation. The timing parameters of the $\Delta B_0$ used for simulation were as follows: $t_{d1}$ = 9 ms, $t_{ru}$ = $t_{rd}$ = 1 ms, $t_{d2}$ = 9.78 ms and $t_{evol}$ was increased in 0.01 ms steps to span the desired simulation range (up to 3000 ms). In a next step, the simulated relaxation curves were compared to the measured curves. All curves were normalized to account for different equilibrium magnetizations and effective offset fields as proposed in [4]. Furthermore, the difference between $M_z^+$ and $M_z^-$ relaxation curves was calculated resulting in the corresponding dreMR signal.

### DreMR imaging

Images at $B_0+\Delta B_0$ (2.99 T) and $B_0-\Delta B_0$ (2.79 T) were acquired to demonstrate the feasibility of generating contrast based only on dispersive properties of the contrast agent. The imaging parameters of the modified SR-SE sequence were as follows: $T_{evol}$ = 150 ms, TR/TE = 3000/15 ms, FOV = 40 mm x 40 mm, matrix size = 192 x 192, BW = 130 Hz/Px, flip angle = 90° and slice thickness = 5 mm. In order to maximize the contrast in the dreMR image the evolution time $T_{evol}$ should be chosen to match approximately the sample's $T_1$ [4]. Again, all images were normalized prior to magnitude subtraction obtaining the final dreMR image.

## 3. Results and Discussion

### Hardware setup

Figure 2a shows ghosting artefacts which arose when the GPA was left enabled during image acquisition. They appeared only in phase encoding direction due to the random phase fluctuations caused by the output current noise of the GPA. The GPA blanking eliminates all visible ghosting artefacts and a well-resolved image of the phantom can be obtained (Figure 2b). In addition to the intrinsic timing restrictions, i.e., 20 ms for enable and 2 ms for disable, the GPA noise might interfere with the magnetization preparation module if an overlap with the RF pulse occurs. Hence, the beginning of the trigger signal must be timed accordingly to prevent such interferences. Figure 3 shows the measured timing between trigger and begin of the field-shift including the onset of increased output noise once the GPA is enabled.

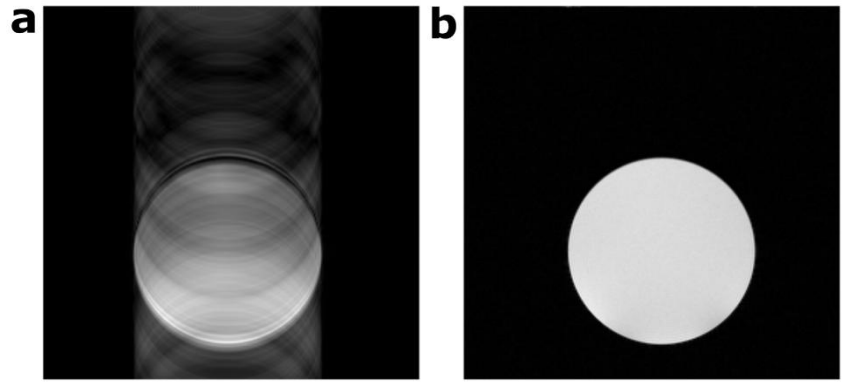

Figure 2: (a) Random phase fluctuations induced by the output current noise of the GPA cause severe ghosting artefacts in phase encoding direction. (b) Disabling the GPA during image acquisition eliminates all visible artefacts.

This required delay time was included in equation (1) as $t_{d1}$ to account for a relaxation with $B_0$ prior the actual begin of the field shift. The GPA blanking approach is rather easy to implement as it utilizes the available control input of the GPA and requires only one additional TTL signal. Of course, the implementation is restricted to amplifiers providing such a dedicated control input and sufficient short enable and disable times. The inherent timing delays can be considered during the design and programming of the FFC-MRI pulse sequence and are not a limiting factor for the FFC-MRI sequences discussed here as the TR is usually much longer than the enable time. However, for other types of sequences, e.g. with very short repetition times, i.e. in the range of the enable time (i.e. 20 ms), other correction methods like in [5,17] must be used.

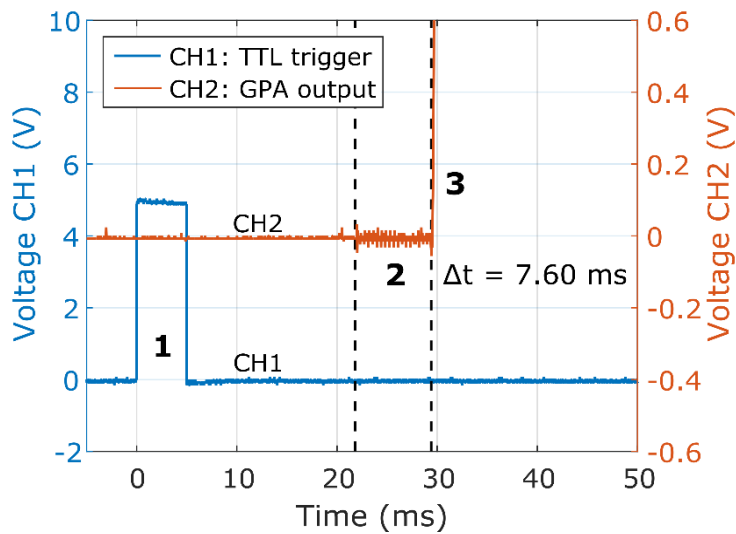

Figure 3: Measured timing of the TTL trigger signal (CH1, lower trace) and the voltage proportional to the output current of the GPA (CH2, upper trace). The rising edge of the $B_0$ waveform (3) occurs 24.5 ms after the end of the trigger pulse (1). The cursors indicate the interval (2) between GPA enable and the rising $B_0$ edge. The increased noise level in this phase is clearly visible and has a duration of Δt = 7.60 ms.

The imaging region of the presented $B_0$ insert coil is limited to phantom and small animal applications. Significant effort has gone into the construction of a fully commissioned human-scale FFC-MRI system with a maximum field strength of 0.2 T [21]. The inherently present $R_1$ dispersion in this field cycling range may have applications in the early diagnosis of a range of disease [22,23]. For dreMR imaging there exist feasibility studies of $B_0$ insert coils designed for human head and prostate imaging [24,25], which is an important step towards clinical applications at high field.

Eddy current characterization and compensation

The measured frequency shift induced by the eddy currents after a positive $\Delta B_0$ pulse is shown in Figure 4a for an amplitude of 100 mT, ramp times of 1ms and an evolution time of 300 ms. Fitting of a mono-exponential model resulted in a model amplitude of 367.6 Hz and a time constant of 169.3 ms with 95% confidence intervals (CI) of [359.4 Hz, 375.8 Hz] and [164.2 ms, 174.9 ms], respectively. Fitting of a bi-exponential model resulted in a slow and fast decaying component with amplitudes of 336.2 Hz and 97.6 Hz and with time constants of 183.6 ms and 17.8 ms, respectively. The CIs were [331.0 Hz, 341.5 Hz] and [87.5 Hz, 107.8 Hz] for the amplitudes and [180.9 ms, 186.4 ms] and [14.8 ms, 22.4 ms] for the time constants, respectively. The bi-exponential model with an adjusted $R^2$ value of 0.9996 performs better than the mono-exponential model with an adjusted $R^2$ value of 0.9951. The better fit can be seen notably at the beginning of the eddy current decay curve in Figure 4a, i.e. in the time interval where the eddy current characterization is crucial for the eDREAM compensation. Given this more accurate description in the beginning of the eddy current decay the bi-exponential model was used for the eDREAM implementation. The uncompensated eddy current effects lead to a rigid-body image shift in frequency and slice encoding direction causing severe artefacts when calculating the difference between reference image and shifted image (Figure 4b). The eddy current artefacts are completely removed by using the eDREAM compensation with the bi-exponential model (Figure 4c). As Gadovist® (Bayer Vital GmbH, Germany) has a negligible magnetic field dependence [26] in the achievable $B_0$ field-cycling range, no significant signal is expected in the difference image.

It is important to note that the dynamics of the eddy current field depend on several parameters of the $\Delta B_0$ pulse such as amplitude, ramp up time, flat top time, ramp down time and repetition time. The results presented herein are exemplary for a field shift of 100 mT, equal ramp times of 1ms each, a $T_{evol}$ of 300 ms and a TR of 3000 ms. For different timing parameters a new characterization becomes necessary. It is to be expected that increasing the ramp times would lead to a decrease of the frequency shift due to a smaller dB/dt and hence lower eddy currents. Furthermore, a TR of 3000 ms is long enough (> 5 time constants) to allow for a sufficient eddy current decay before applying the next $\Delta B_0$ pulse as can be seen in Figure 4a where the measured frequency shift has decayed below 3 Hz after 900 ms. Short TRs (< 5 time constants) lead to a formation of a dynamic steady state as the eddy current fields of the previous $\Delta B_0$ pulse add to the field induced by the next $\Delta B_0$ pulse. A comprehensive analysis of eddy current formation and the eDREAM compensation has been given by Hoelscher et al. in [7].

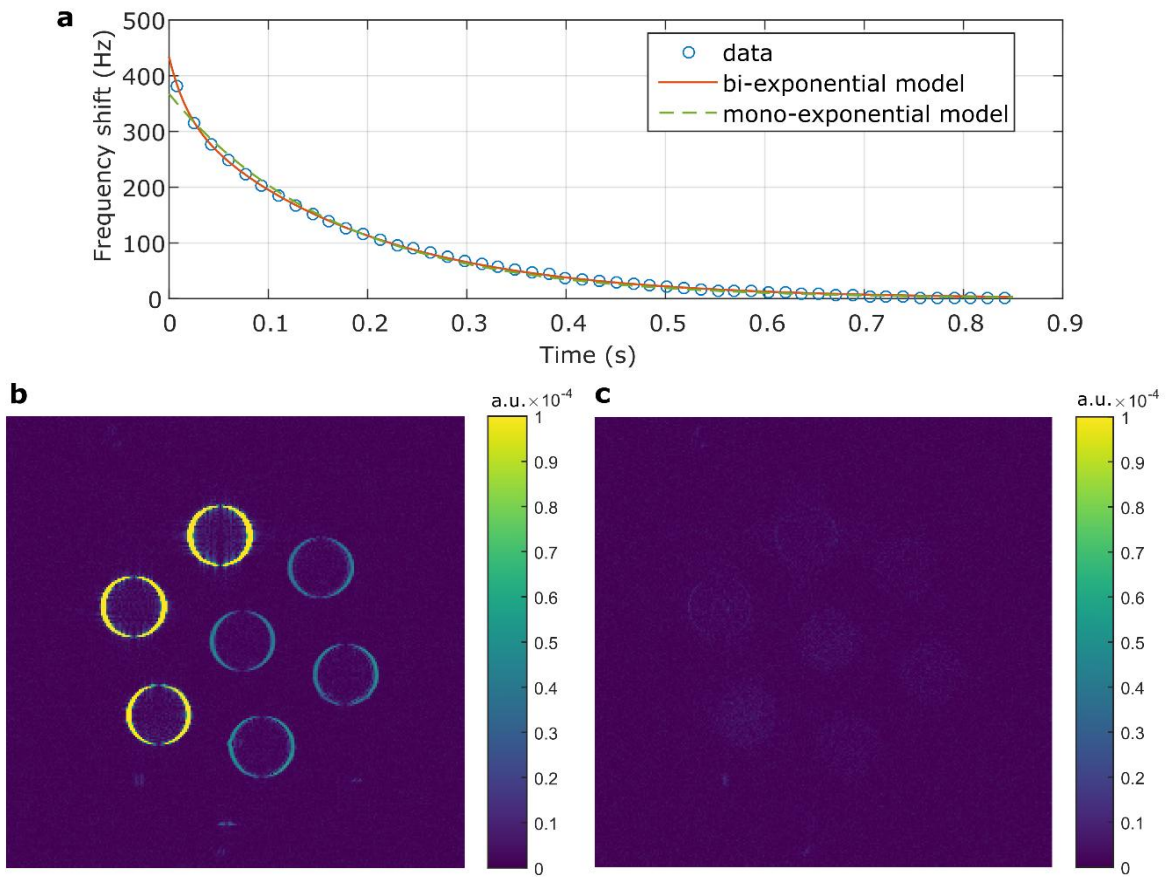

Figure 4: (a) Measured frequency shift induced by the eddy currents (blue circles) for a 100 mT ΔB$_0$ pulse with T$_{evol}$ of 300 ms and a ramp times of 1ms. A mono-exponential (green line) as well as a bi-exponential (red line) model were fitted to the data. This characterization of the eddy current field dynamics allows for a compensation by means of dynamic reference phase modulation. (b) Difference between reference image and image at shifted B$_0$ without eddy current compensation and (c) with compensation using the bi-exponential model.

Dispersive contrast agent for 3T and FFC measurements

Both the cube-shaped as well as the spherically shaped IOMNPs proved to be suitable for system validation at a field strength of 3 T as they exhibit a pronounced peak in the NMRD profile (Figure 5) around 20 MHz followed by a steep downward slope in the targeted frequency range of 119-127.4 MHz corresponding to 2.89T±100 mT. In contrast, the hexane sample shows only a weak dependence on the magnetic field i.e. a flat slope in the NMRD profile within the achievable field-cycling range and is therefore suited as a non-dispersive reference.

Furthermore, for visual comparison of all samples, R$_1$ maps for 2.79 T and 2.99 T together with the corresponding R$_1$ dispersion map are shown in Figure 6b-d, respectively. The spherical IOMNPs (C8_hex) show higher R$_1$ values and a steeper slope of the NMRD profile (Figure 5) in comparison to the cubic IOMNPs (S8_hex). This is also emphasized by the R$_1$ dispersion map in Figure 6d, giving a good linear approximation of the slope in the NMRD profile within the achievable field-cycling range i.e. $\Delta R_1/\Delta B_0$. As expected, no significant $\Delta R_1/\Delta B_0$ is visible in the $R_1$ dispersion map for the hexane reference sample.

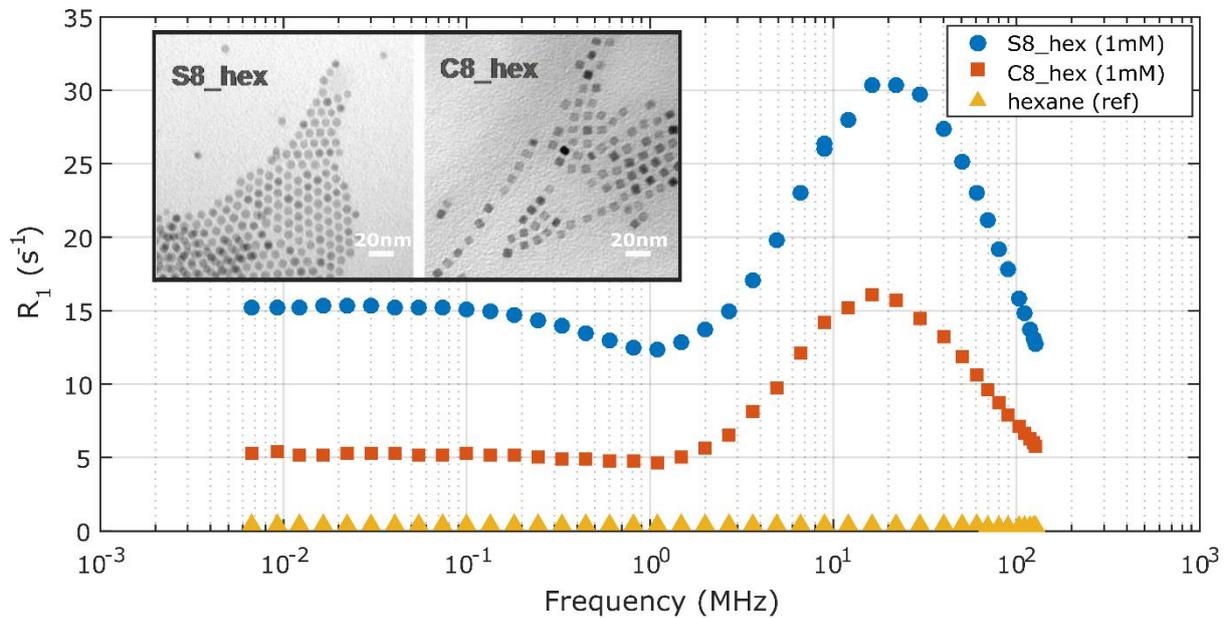

Figure 5: $^1$H $R_1$ NMRD profiles of the cubic (C8_hex, red squares) and spherical (S8_hex, blue circles) IOMNPs with 1 mM concentration as well as of pure hexane (ref, yellow triangles) measured up to a field strength of 3 T (127.8 MHz) using a commercial FFC-NMR relaxometer. All measurements were performed at room temperature (295 K). The insets show representative TEM images of S8 and C8 samples, indicating that the used IOMNPs have different morphology and are nearly monodisperse in size and shape.

Table 1 compares the $R_1$ values obtained with FFC-MRI measurements and the corresponding data points in the NMRD profile (2.79 T, 2.89 T and 2.99 T) for C8_hex and S8_hex IOMNPs samples with 1 mM concentration. In order to exclude pixels at the edges from the evaluation, the $R_1$ values have been estimated over a rectangular ROI in the middle of sample 1 and 4 in Figure 6a. The $R_1$ estimation with the proposed FFC-MRI system is in accordance with the FFC-NMR measurements (Stelar, Spinmaster). Compared to the FFC-NMR measurements, the $R_1$ values obtained with FFC-MRI are slightly lower for the S8_hex samples and slightly higher for the C8_hex samples with mean relative errors of -11.6 % and +7.6 %, respectively. It is likely that this systematic error results from different sample concentrations due to inaccuracies in the sample preparation and possible sample degradation during delivery to the FFC-NMR facility. Regardless of this error, the agreement between FFC-MRI and FFC-NMR measurements is sufficient for system validation.

All used IOMNPs were dispersed in hexane as they are produced by a synthetic route which leads to hydrophobic nanocrystals with high crystallinity and highest size and shape control achievable to date. The inset in Figure 5 reports representative TEM images of the C8 and S8 IOMNPs, showing that both samples are nearly monodisperse in size and shape. The use of the IOMNPs dispersed in an organic solvent is therefore justified for the validation of the proposed FFC-MRI system, because the signal-to-noise ratio (SNR) is important for an accurate characterization. Theoretically, it is possible to dissolve them in an aqueous solution, but this comes at the expense of losing $R_1$ dispersion [19]. The development of strategies for the dispersion in water without losing the beneficial relaxometric properties and achievable SNR, should be investigated in a separate study to enable direct applicability of the IOMNPs to in-vivo experiments.

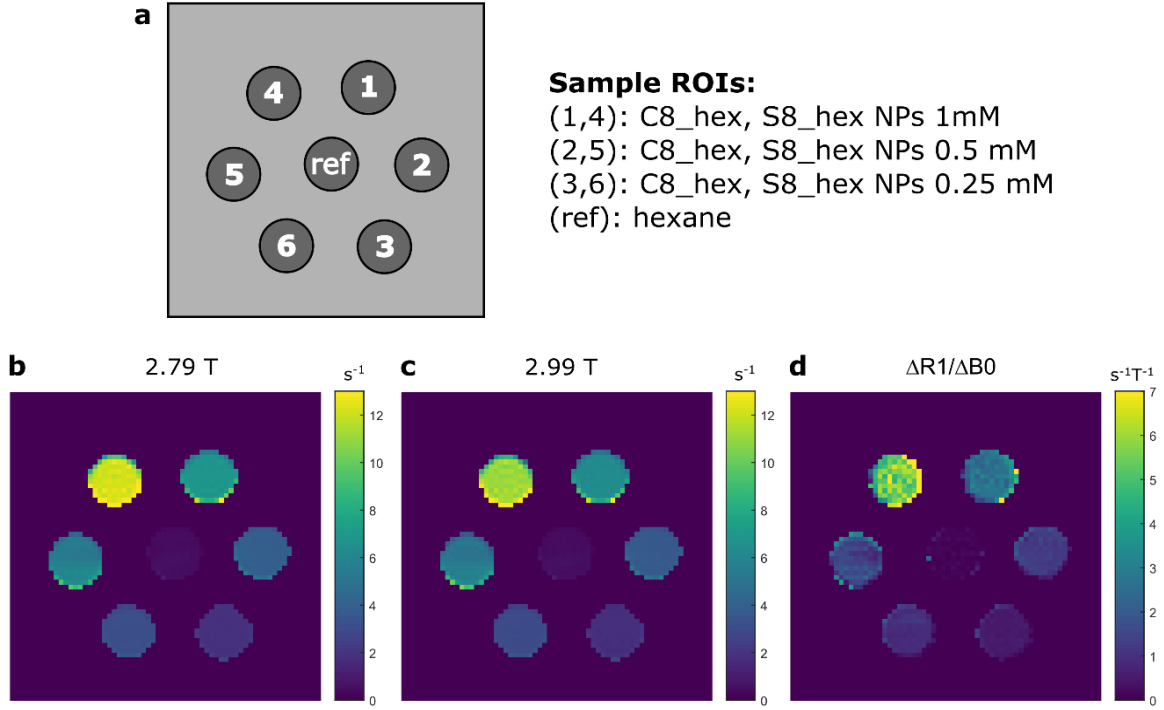

Figure 6: (a) Sample arrangement in the phantom containing three different concentrations of cubic (C8_hex) and spherical (S8_hex) IOMNPs as well as a non-dispersive hexane reference sample (ref). (b) and (c) show $R_1$ maps calculated on a pixel-by-pixel basis for a field strength of 2.79T and 2.99T, respectively. (d) $R_1$ dispersion map $\Delta R_1/\Delta B_0$ (magnitude) obtained from (b) and (c).

Table 1: Comparison of $R_1$ at different field strengths measured with a commercial FFC-NMR relaxometer and with the proposed FFC-MRI system (mean ± standard deviation). For the FFC-MRI method, $R_1$ values were obtained by ROI analysis (rectangular ROI in the middle of each sample).

| Sample | Method | Longitudinal relaxation rates ($s^{-1}$) | | |
| --- | --- | --- | --- | --- |
| | | 2.79T | 2.89T | 2.99T |
| S8_hex (1mM) | FFC-NMR | 13.62 (±0.14) | 13.17 (±0.13) | 12.77 (±0.13) |
| | FFC-MRI | 12.21 (±0.31) | 11.63 (±0.34) | 11.14 (±0.35) |
| C8_hex (1mM) | FFC-NMR | 6.25 (±0.06) | 6.07 (±0.06) | 5.93 (±0.06) |
| | FFC-MRI | 6.81 (±0.13) | 6.54 (±0.14) | 6.29 (±0.13) |

DreMR signal simulation

The behaviour of the longitudinal magnetization $M_z$ in the presence of a trapezoidal field shift applied during the evolution phase was simulated using equation (1). Simulated relaxation curves (normalized to $M_0$) for a positive and negative field shift of 100 mT are shown in Figure 7a and compared to FFC-MRI measurement data for various evolution times ($T_{evol}$ = 60, 100,

150, 300, 500, 800, 1500 and 3000 ms). The measured relaxation curves show a high degree of congruence with the simulation, which accounts for all underlying timing restrictions in the $\Delta B_0$ pulse. The applied field shifts of ±100mT result in a relaxation to different equilibrium magnetizations of $M_0^\pm = M_0 \pm 3.3\%$. Theoretically, a relaxation to $M_0^\pm = M_0 \pm 3.46\%$ ($M_0(1\pm\Delta B_0/B_0)$) is expected for a rectangular field shift without timing restrictions such as finite ramp times and delay times $t_{d1}$ and $t_{d2}$. In order to isolate only the effect of $R_1$ dispersion in the dreMR signal a proper field-dependent scaling is of great importance as discussed by Hoelscher et al. in references [4,27]. Figure 7b shows the characteristic dreMR signal (magnitude) obtained after subtraction of the relaxation curves for 2.79 T and 2.99T. Prior to subtraction the relaxation curves were normalized to account for relaxation to different $M_0$ and effective offset fields. The measured data is in good agreement with the dreMR signal simulation after accounting for all timing restrictions. Furthermore, we observe a minimum (for negative $\Delta R_1/\Delta B_0$) located around the $T_1$ of the sample (ROI 2) which is approximately 267 ms. As dreMR images are based on image subtraction they suffer from a poor SNR. In order to maximize the dreMR contrast for a specific sample the evolution time $T_{evol}$ should be chosen to match the sample's $T_1$ [4]. Therefore, equation (1) can be used to optimize dreMR imaging parameters under consideration of all timing restrictions of the $\Delta B_0$ pulse in the MRI pulse sequence.

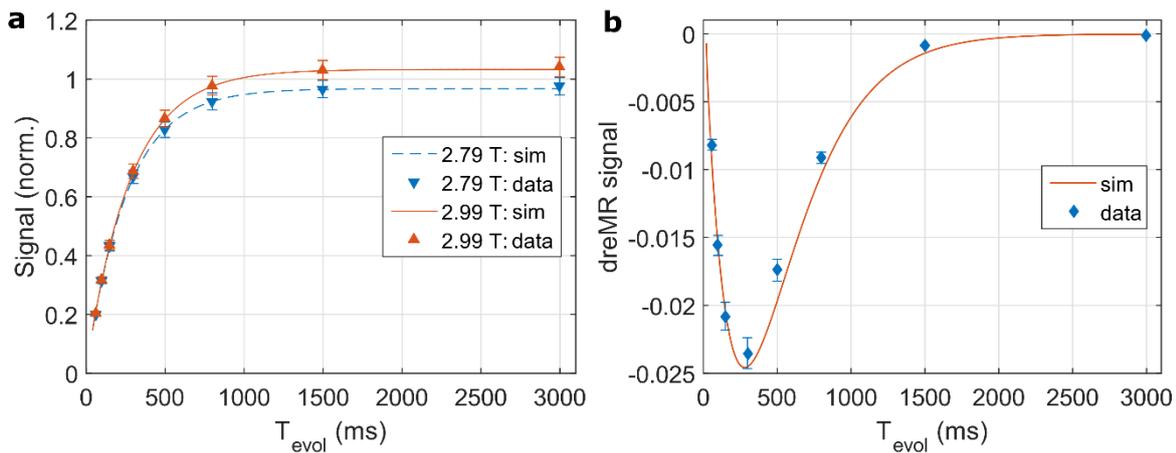

Figure 7: (a) Simulated longitudinal relaxation curves for 2.99 T (red solid line) and 2.79 T (blue dashed line) compared to relaxation curves measured (mean ± standard deviation of ROI 2 in Figure 6a) with the proposed FFC-MRI system (upward-pointing triangle for 2.99 T and downward-pointing triangle for 2.79T). (b) Simulated and measured dreMR signal obtained by subtraction of the relaxation curves in (a) after normalization to account for different equilibrium magnetizations and effective $\Delta B_0$ offset fields.

DreMR imaging

Figure 8 shows a dreMR image acquired with a $T_{evol}$ of 150 ms and an isotropic in-plane resolution of 0.2 mm. As can be clearly seen, signal arises only from samples containing the dispersive IOMNPs, whereas the signal is strongly suppressed in the hexane reference sample. When comparing samples containing cubic shaped IOMNPS (samples 1-3 in Figure 6a), theoretically, maximum signal intensity should be observed for sample 1 as $T_{evol} \approx T_1$. In practice, the dreMR contrast in Figure 8 is altered by an additional $T_2$-weighting due to short $T_2$ of the IOMNPs [19]. Consequently, the samples with the highest concentrations seemingly paradoxically, show less contrast enhancement, though, in reality, their $T_1$ enhancement is maximal. Nevertheless, this dreMR image is a clear proof-of-principle, demonstrating the capability of generating $R_1$ dispersion contrast with suppressed background signal at a clinical field strength of 3 T. This proof is sufficient for the purpose of system validation only. For improved evaluation of signal contrasts, the TE of 15 ms has to be reduced to the order of a few ms to minimize the $T_2$-weighting, which was not feasible with the modified SR-SE sequence used herein.

Theoretically, a twofold increase in SNR of the images prior dreMR image subtraction is to be expected for 3 T compared to 1.5 T [28]. Aside from this potential benefit, the SNR increase in the dreMR image may be compensated by a reduced $R_1$ dispersion at 3 T. Whether a field strength of 1.5 T or 3 T is more beneficial for dreMR imaging depends strongly on the NMRD profile of the contrast agent in use. In the context of prospective QRE based contrast agents, it is favourable to have a wide variety of field-cycling ranges as it is completely unexplored at which field strength such compounds give the best image contrast. Therefore, the presented FFC-MRI system expands the field cycling technique with an additional cycling range centered at 2.89 T.

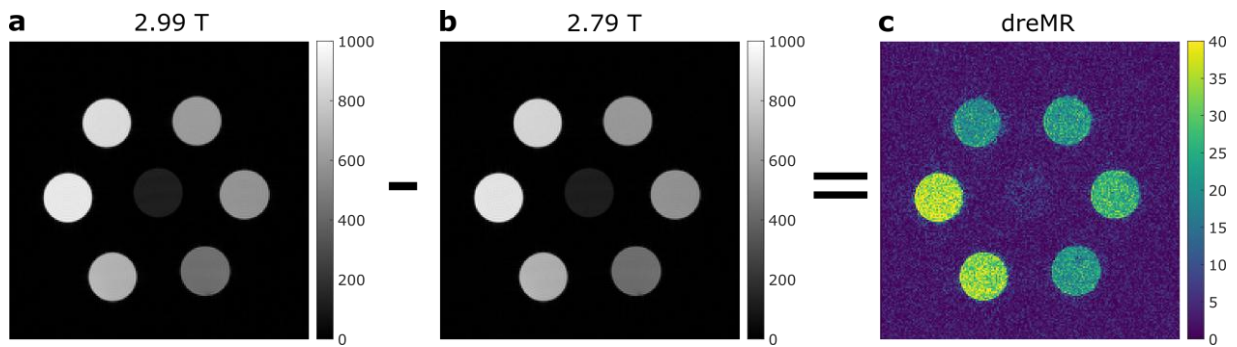

Figure 8: Images acquired for a field strength of 2.99 T (a) and 2.79 T (b) with an isotropic in-plane resolution of 0.2 mm and $T_{evol}$ of 150 ms were normalized to account for different $M_0$ and effective offset fields. Image subtraction of (a) and (b) resulted in the corresponding dreMR image whereof the magnitude is shown in (c).

## 4. Conclusions

In conclusion, we have successfully implemented a small animal FFC-MRI system for a clinical field strength of 3 T by inserting an additional $B_0$ insert coil into an otherwise conventional MRI system. Iron oxide magnetic nanoparticles proofed to be suitable for system validation as they exhibit a sufficiently high $R_1$ relaxation dispersion in the achievable field-cycling range of ±100 mT around the nominal field strength of 2.89 T. System validation was successfully performed by comparison of FFC-MRI measurements with FFC-NMR relaxometry. Furthermore, a proof-of-principle for dreMR imaging at 3 T has been achieved by generating contrast arising only from the dispersive properties of the contrast agent in use. In the context of developing new contrast agents, this hardware implementation and validation provides a ready-to-use hardware setup for investigating dispersive properties of MRI contrast agents at a field strength of 3 T.

# Acknowledgement

This project receives financial support by the European Commission in the frame of the H2020 Programme (FET-open) under grant agreement 665172. This article is also partially based upon work from COST Action CA15209, supported by COST (European Cooperation in Science and Technology). The authors would like to thank Prof. A. Lascialfari for discussion within the framework of COST Action CA15209.